\documentclass[aip,rsi,superscriptaddress,amsmath,amssymb,reprint,longbibliography]{revtex4-2}
\usepackage{graphicx}
\usepackage{lettrine}
\usepackage{adjustbox}
\usepackage{multirow}

\begin{document}
\title{Imaging the decay of quantized vortex rings to decipher quantum dissipation}
\author{Yuan Tang}
\affiliation{National High Magnetic Field Laboratory, 1800 East Paul Dirac Drive, Tallahassee, FL 32310, USA}
\affiliation{Mechanical Engineering Department, FAMU-FSU College of Engineering, Florida State University, Tallahassee, FL 32310, USA}

\author{Wei Guo}
\email[Email: ]{wguo@magnet.fsu.edu}
\affiliation{National High Magnetic Field Laboratory, 1800 East Paul Dirac Drive, Tallahassee, FL 32310, USA}
\affiliation{Mechanical Engineering Department, FAMU-FSU College of Engineering, Florida State University, Tallahassee, FL 32310, USA}

\author{Hiromichi Kobayashi}
\affiliation{Research and Education Center for Natural Sciences, Keio University, 4-1-1 Hiyoshi, Kohoku-ku, Yokohama 223-8521, Japan}
\affiliation{Department of Physics, Hiyoshi Campus, Keio University, 4-1-1 Hiyoshi, Kohoku-ku, Yokohama 223-8521, Japan}

\author{Satoshi Yui}
\affiliation{Department of Physics, Osaka Metropolitan University, 3-3-138 Sugimoto, Sumiyoshi-ku, Osaka 558-8585, Japan}
\affiliation{Nambu Yoichiro Institute of Theoretical and Experimental Physics (NITEP), Osaka Metropolitan University, Osaka 558-8585, Japan}

\author{Makoto Tsubota}
\affiliation{Department of Physics, Osaka Metropolitan University, 3-3-138 Sugimoto, Sumiyoshi-ku, Osaka 558-8585, Japan}
\affiliation{Nambu Yoichiro Institute of Theoretical and Experimental Physics (NITEP), Osaka Metropolitan University, Osaka 558-8585, Japan}

\author{Toshiaki Kanai}
\affiliation{National High Magnetic Field Laboratory, 1800 East Paul Dirac Drive, Tallahassee, FL 32310, USA}
\affiliation{Department of Physics, Florida State University, Tallahassee, FL 32306, USA}


\begin{abstract}
Like many quantum fluids, superfluid helium-4 (He II) can be considered as a mixture of two miscible fluid components: an inviscid superfluid and a viscous normal fluid consisting of thermal quasiparticles~\cite{Landau-book}. A mutual friction between the two fluids can emerge due to quasiparticles scattering off quantized vortex lines in the superfluid~\cite{Vinen-1957-PRS-III}. This quantum dissipation mechanism is the key for understanding various fascinating behaviors of the two-fluid system~\cite{Tisza-1938-Nature,VanSciver-2012-book}. However, due to the lack of experimental data for guidance, modeling the mutual friction between individual vortices and the normal fluid remains an unsettled topic despite decades of research~\cite{Schwarz-1977-PRL,Schwarz-1988-PRB,Idowu-2000-PRB,Kivotides-2018-PRF,Galantucci-2020-EPJ,Yui-2020-PRL}. Here we report an experiment where we visualize the motion of quantized vortex rings in He II by decorating them with solidified deuterium tracer particles. By examining how the rings spontaneously shrink and accelerate, we provide unequivocal evidences showing that only a recent theory~\cite{Galantucci-2020-EPJ} which accounts for the coupled motion of the two fluids with a self-consistent local friction can reproduce the observed ring dynamics. Our work eliminates long-standing ambiguities in our theoretical description of the vortex dynamics in He II, which will have a far-reaching impact since similar mutual friction concept has been adopted for a wide variety of quantum two-fluid systems, including atomic Bose-Einstein condensates (BECs)~\cite{Haljan-2001-PRL,Kwon-2021-Nature}, superfluid neutron stars~\cite{Greenstein-1970-Nature,Packard-1972-PRL,Andersson-2006-MNRAS}, and gravity-mapped holographic superfluid~\cite{Chesler-2013-Science,Wittmer-2021-PRL}.
\end{abstract}
\maketitle

Quantized vortices are topological defects in the superfluid. In 3D space, they appear as density-depleted thin tubes (e.g., tube core radius $\sim$1~{\AA} in He II~\cite{Donnelly-1991-B}), each carrying a circulating flow with a fixed circulation $\kappa=h/m$, where $h$ is Planck's constant and $m$ is the mass of the bosons constituting the superfluid~\cite{Tilley-1990-book}. The motion of the quantized vortices is responsible for a wide range of phenomena in diverse quantum-fluid systems, such as the emergence of quantum turbulence in He II and atomic BECs~\cite{Vinen-2007-book,Tsubota-2013-book}, the initiation of dissipation in type-II superconductors~\cite{Larbalestier-2001-Nature}, the appearance of glitches in neutron star rotation~\cite{Greenstein-1970-Nature,Packard-1972-PRL}, and the formation of cosmic-string network~\cite{Zurek-1985-Nature}. Developing a theoretical model to reliably predict the vortex motion in quantum fluids in the presence of the thermal component promises a broad significance spanning multiple physical science disciplines.

\renewcommand{\figurename}{\textbf{Fig.}}
\begin{figure*}[t]
\centering
\includegraphics[width=1\linewidth]{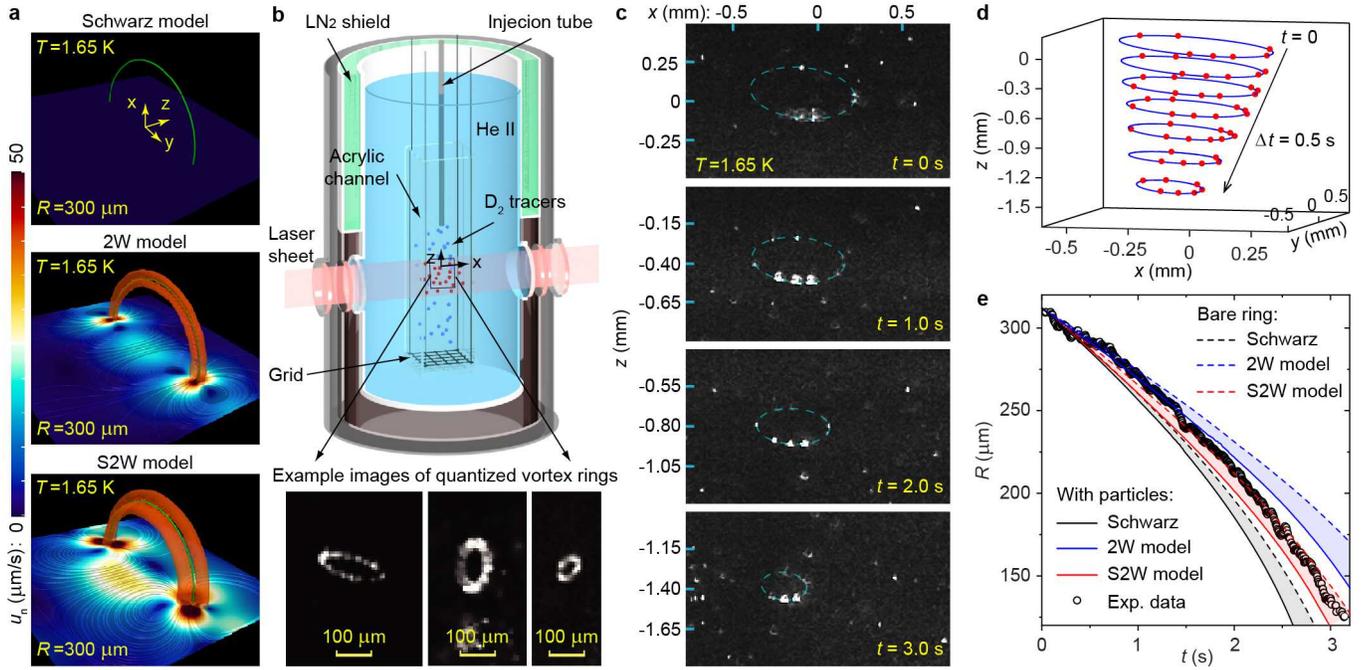}
\caption{\textbf{Modeling and imaging quantized vortex rings in He II.} \textbf{a}, Calculated normal-fluid velocity field $\mathbf{u}_n$ around a quantized vortex ring in He II. Due to the axial symmetry, we only show $\mathbf{u}_n$ in the $y$-$z$ plane and the vortex ring above the plane (i.e., the green curve). The normal-fluid vortex rings (reddish half circles) are rendered in the same way as in Ref.~\cite{Yui-2020-PRL}. \textbf{b}, Schematic diagram of the experimental setup. \textbf{c}, Images showing the D$_2$ particles (white dots) trapped on a moving vortex ring in quiescent He II. The dashed ellipse is a fit to the trapped particles' positions. \textbf{d}, Obtained vortex-ring profile with the trapped particles (red dots) at different times. \textbf{e}, Comparison of the observed ring radius $R(t)$ with model simulations.}
\label{Fig1}
\end{figure*}

In the pioneering work of Schwarz~\cite{Schwarz-1977-PRL,Schwarz-1988-PRB}, a vortex filament model was developed for studying turbulence in He II. In this model, the quantized vortices are described by zero-thickness filaments that are divided into small segments. A vortex segment with a length $\Delta \xi$ located at $\mathbf{s}$ would experience a Magnus force $\mathbf{f}_M=\rho_s\kappa \mathbf{s}'\times(\mathbf{u}_L-\mathbf{u}_s)\Delta \xi$ when its velocity $\mathbf{u}_L$ differs from the local superfluid velocity $\mathbf{u}_s$. Here $\mathbf{s}'$ is the unit tangent vector along the filament, and $\rho_s$ is the superfluid density. Furthermore, any relative motion between the vortex segment and the normal fluid would result in a mutual friction force as derived by Schwarz $\mathbf{f}_{sn}=[-\gamma_0\mathbf{s}'\times(\mathbf{s}'\times(\mathbf{u}_n-\mathbf{u}_L))+\gamma_0'\mathbf{s}'\times(\mathbf{u}_n-\mathbf{u}_L)]\Delta \xi$, where $\gamma_0$ and $\gamma_0'$ are temperature-dependent empirical coefficients~\cite{Schwarz-1988-PRB}. By balancing the two forces, Schwarz obtained the vortex equation of motion (see Methods), which has been extensively employed in past studies of vortex dynamics~\cite{Adachi-2010-PRB,Risto-2014-PNAS,Yui-2022-PRL}.

However, a known limitation of the Schwarz model is that the normal-fluid velocity $\mathbf{u}_n$ is prescribed and there is no back action from the vortices to the normal fluid. To fix this issue, a two-way (2W) model was later developed, where $\mathbf{u}_n$ is solved using the Navier-Stokes equation with an added mutual-friction term that couples to the vortices. This model has allowed researchers to explain puzzling observations in He II turbulence~\cite{Yui-2018-PRL,Yui-2020-PRL}. Nonetheless, it was postulated that the coefficients $\gamma_0$ and $\gamma_0'$ may not be applicable to individual vortices since they were deduced from measurements where $\mathbf{u}_n$ was averaged over an array of vortices~\cite{Idowu-2000-PRB}. Over the past two decades, researchers have strived to calculate the friction coefficients in a self-consistent manner~\cite{Idowu-2000-PRB,Kivotides-2018-PRF,Galantucci-2020-EPJ}. These efforts led to the striking prediction of the triple-vortex-ring structure in He II~\cite{Idowu-2000-PRB,Kivotides-2000-Science}. Recently, Galantucci \emph{et al.} derived the most refined version of the self-consistent two-way (S2W) model where the mutual friction coefficient can be calculated directly from $\mathbf{u}_n$ without any empirical experimental input~\cite{Galantucci-2020-EPJ}.

These different models render distinct normal-fluid flow structures around the quantized vortices, which affect the vortex motion. As an example, we show in Fig.~\ref{Fig1}a the calculated normal-fluid velocity field around a quantized vortex ring in quiescent He II using all three models (see Methods for details). Unlike the Schwarz model where $\mathbf{u}_n=0$, both the 2W model and the S2W model reveal two oppositely polarized normal-fluid vortex rings sandwiching the quantized vortex ring. These normal-fluid rings affect the local $\mathbf{u}_n$ experienced by the quantized ring and hence can alter the mutual friction dissipation. However, is this triple-ring structure real? If so, which model better describes the true vortex dynamics? These questions are important but have remained opened due to the lack of experimental information. In this work, we provide the long-awaited data to show that only the S2W model can reproduce experimental observations. This decisive study will break new ground for modeling and understanding various vortex-involved phenomena in quantum two-fluid systems.

\renewcommand{\figurename}{\textbf{Fig.}}
\begin{figure*}[t]
\centering
\includegraphics[width=1\linewidth]{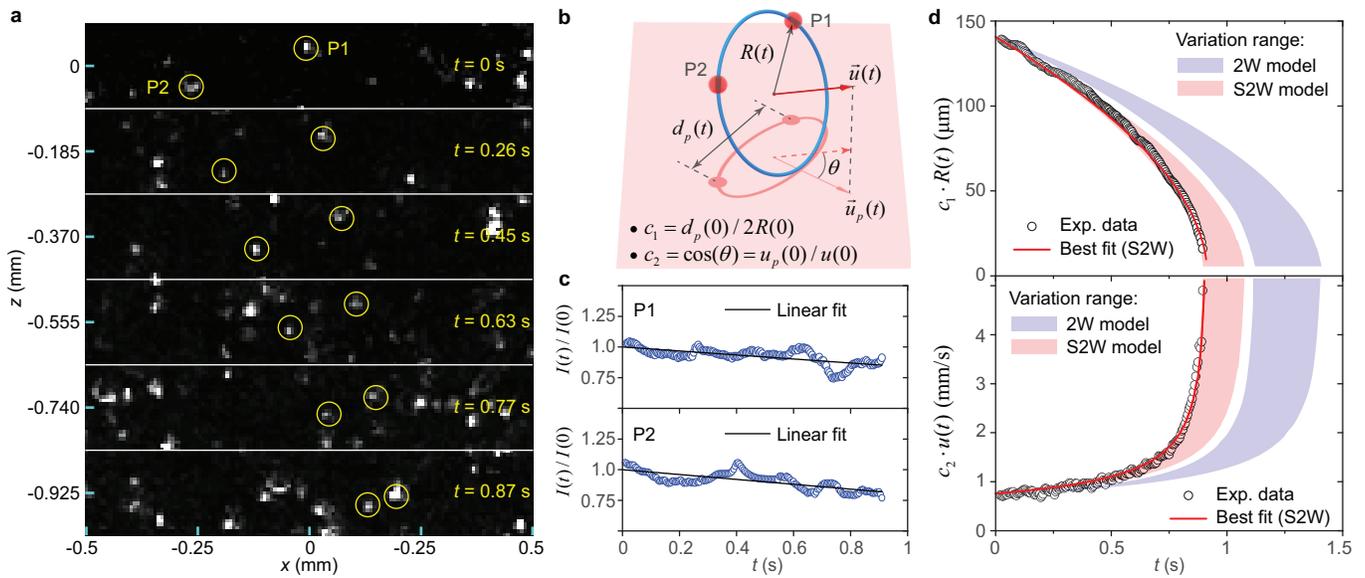}
\caption{\textbf{Analysis of a vortex ring with two trapped particles.} \textbf{a}, Images showing two trapped particles (circled) moving in He II at 1.65~K. \textbf{b}, A schematic explaining the concept of the projection parameters $c_1$ and $c_2$. \textbf{c}, Variation of the brightness of the two trapped particles. \textbf{d}, Comparison of the projected ring radius and velocity data with model simulations.
}
\label{Fig2}
\end{figure*}

\section*{Visualizing quantized vortex rings}\label{SecII}
To study the vortex motion, we visualize quantized vortices in He II by decorating them with solidified deuterium (D$_2$) tracer particles~\cite{Bewley-2006-Nature,Guo-2014-PNAS}. This method has already allowed researchers to gain valuable insights into the properties of tangled vortices~\cite{La_Mantia-2014-EL,Mastracci-2019-PRF,Fonda-2019-PNAS,Tang-2021-PNAS}. However, past attempts to image vortex rings failed to produce useful data mainly due to two issues~\cite{Bewley-2009-JLTP}: 1) vortex-ring events were scarce because of the low vortex-line density in the experiment; and 2) too many particles condensed on the vortex cores which altered the core size and hence the ring dynamics. To fix these issues, we control the vortex generation by towing a mesh grid in a plexiglass channel (1.6$\times$1.6$\times$33~cm$^3$) in He II (see Fig.~\ref{Fig1}b). Following the grid motion, a mixture of D$_2$ gas and $^4$He gas is injected into the channel at about 30~s delay so that the background flow is weak but vortices with a line density of the order 10$^2$~cm$^{-2}$ still remain~\cite{Mastracci-2018-RSI,Tang-2020-PRF}. The D$_2$ gas forms ice particles with a mean radius of 1.1~$\mu$m as determined from their settling velocities (see Methods). When the D$_2$ particles are close to the vortex cores, they get trapped on the vortices due to a Bernoulli pressure caused by the circulating superfluid~\cite{Donnelly-1991-B}. Through extensive trials, we have figured out the optimal injection parameters to achieve the desired particle number density on the vortices. The particles are then illuminated by a laser sheet (thickness 0.8~mm) and their positions are recorded at 200 Hz by a video camera placed perpendicular to the laser plane. Occasionally, we can see vortex rings propagating within the laser sheet. A collection of representative ring events are included in Supplementary Video 1. We have also captured videos showing for the first time how vortex rings are created by reconnections of intersecting vortex lines (see Supplementary Video 2).


\section*{Data analysis and model comparison}\label{SecIII}
To extract useful information on vortex-ring propagation, we focus on analyzing selected events where the rings are decorated by discrete D$_2$ particles and move in He II with negligible background flows. A good example is shown in Fig.~\ref{Fig1}c where the ring moves downward carrying nine D$_2$ particles (see Supplementary Video 3). We first use a feature-point tracking routine~\cite{Sbalzarini-2005-JSB} to determine the positions of the trapped particles in each image. Then, the particle positions are fitted with an ellipse. This fitting, which requires at least 5 particles on the ring, allows us to determine both the ring radius $R$ and the orientation of the ring plane (see Methods). Fig.~\ref{Fig1}d shows the extracted ring profile with the trapped particles at different times. The ring shrinks due to the mutual friction dissipation, which leads to an acceleration of its self-induced motion~\cite{Donnelly-1991-B}. Interestingly, we find that the trapped particles do not move along the vortex core, which may support the core-damping idea proposed by Skoblin \emph{et al.}~\cite{Skoblin-2020-JLTP}. In Fig.~\ref{Fig1}e, we show the obtained $R(t)$ data. For comparison, we also include the simulated $R(t)$ for a bare vortex ring in quiescent He II with the same initial radius using all three models. It appears that the S2W model renders the best agreement with the data.

Nonetheless, the trapped particles can result in additional forces on the vortex core and hence affect the ring's motion~\cite{Mineda-2013-PRB,Barenghi-2009-PRB}. Following Mineda \emph{et al.}~\cite{Mineda-2013-PRB} (see Methods), we consider the Stokes drag~\cite{Landau-book} $\mathbf{f}_D=-6\pi a \mu_n(\mathbf{u}_L-\mathbf{u}_n)$, the gravitational force, and the inertial effect of each trapped particle on the ring. Here $\mu_n$ is the He II dynamic viscosity and $a$ is the particle radius. To evaluate $a$, we first develop a correlation between the particle's brightness $I$ and its radius by comparing the distributions of these two quantities (see Methods). We then examine the time-averaged brightness of each trapped particle and calculate its radius using the correlation. The obtained radiuses are listed in the Extended Data Table~\ref{Tab1}. With this information, we can re-calculated $R(t)$ using the three models (see Fig.~\ref{Fig1}e). Due to the additional Stokes drag, the ring shrinks faster in all three models. Obviously, the Schwarz model overestimates the dissipation and can be rejected. But it becomes less clear whether the S2W model still describes the data better than the 2W model. To make a reliable judgement on these two models, it is imperative to analyze rings with minimal number of trapped particles, since possible uncertainties in the particle size could shift the calculated $R(t)$ curves.

\renewcommand{\figurename}{\textbf{Fig.}}
\begin{figure*}[t]
\centering
\includegraphics[width=1\linewidth]{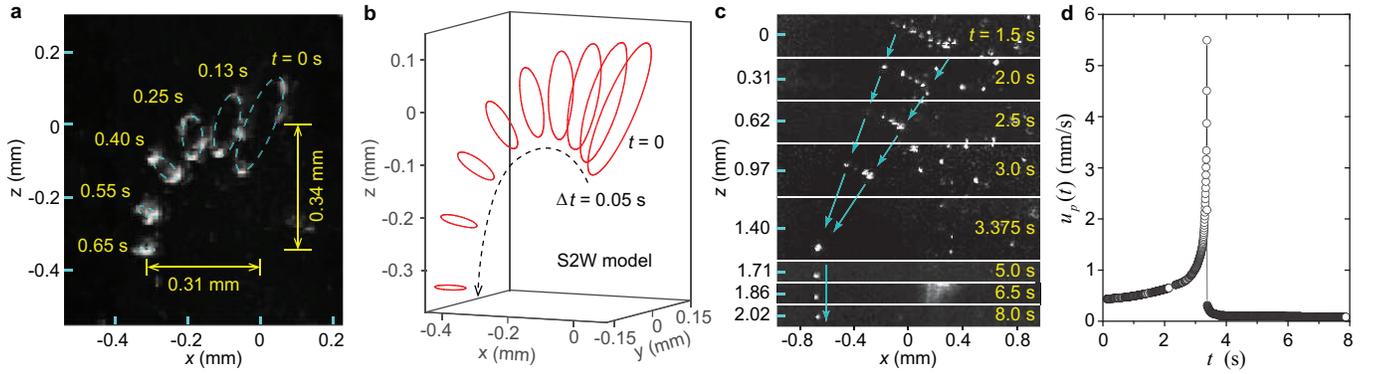}
\caption{\textbf{Other intriguing observations of the vortex rings.} \textbf{a}, A superimposed image showing a heavily doped vortex ring gradually bends its path downward while it shrinks. \textbf{b}, Simulated motion of a vortex ring with the same initial profile carrying 36 D$_2$ particles (4.9 $\mu$m in radius) using the S2W model. \textbf{c}, Images showing how the D$_2$ particles trapped on a vortex ring eventually form a cluster that falls freely in He II. \textbf{d}, Measured centroid velocity of the trapped D$_2$ particles shown in \textbf{c}.
}
\label{Fig3}
\end{figure*}

Luckily, we have recorded several unique events where the rings are decorated by only two D$_2$ particles (see Supplementary Video 4). For these events, the estimated Stokes drag and the gravitational force are only a few percent of the mutual friction. Fig.~\ref{Fig2}a shows our best example, where two particles P$_1$ and P$_2$ move in sync while approaching each other due to the shrinkage of the vortex ring. We can measure the separation distance $d_p(t)$ between the two particles and their centroid velocity $u_p(t)=|\frac{1}{2}(\mathbf{u}_1+\mathbf{u}_2)|$. However, as illustrated in Fig.~\ref{Fig2}b, $d_p(t)$ in general does not equal the vortex-ring diameter $2R(t)$, and $u_p(t)$ can differ from the actual ring velocity $u(t)$ since a projection angle $\theta$ may exist between the ring's propagation direction and the laser plane. In order to utilize the experimental data for model comparison, we adopt the following procedures. First, we assume an initial ring radius $R(0)$ and calculate the evolution of the ring's radius $R(t)$ and velocity $u(t)$ using both the 2W and the S2W models. Next, we evaluate two projection parameters $c_1=d_p(0)/2R(0)$ and $c_2=u_p(0)/u(0)$. These two parameters remain nearly constant because: 1) the particles do not slide along the vortex core as we learned from the study of rings with 5 or more trapped particles; and 2) the centroid of P$_1$ and P$_2$ moves in a straight path, suggesting a constant projection angle. Using $c_1$ and $c_2$, we can then compare $c_1R(t)$ and $c_2u(t)$ directly with the experimental data $d_p(t)/2$ and $u_p(t)$. Finally, we vary $R(0)$ to see which model can render results that simultaneously match $d_p(t)/2$ and $u_p(t)$.

In this analysis, there are a few constraints on the range of $R(0)$ that we can explore. First, $R(0)\geq d_p(0)/2$ since the two particles cannot be separated by more than the diameter of the ring. Second, $u(0)\geq u_p(0)$ due to the projection, which sets an upper limit of $R(0)$ because $u(0)$ drops as $R(0)$ increases. The last constraint comes from the observed particle brightness $I$. As shown in Fig.~\ref{Fig2}c, $I$ for either particles only drops by less than 20\% during the ring's propagation. Based on the cross-sectional profile of the laser sheet (see Methods), we estimate that the ring can move by at most 0.2 mm perpendicular to the laser plane. This sets an upper limit of the projection angle $\theta$, which constrains $u(0)$ and hence $R(0)$. In Fig.~\ref{Fig2}d, we show the calculated $c_1R(t)$ and $c_2u(t)$ using the 2W and the S2W models while $R(0)$ is varied in the range set by all the constraints. Clearly, the experimental data are outside the variation range of the 2W model. On the other hand, we find that the S2W model can nicely reproduce both $d_p(t)/2$ and $u_p(t)$ data at $R(0)=140.8$~$\mu$m. This optimal $R(0)$ is close to $d_p(0)/2=140.6$~$\mu$m, which suggests that the two particles were located nearly across the diameter of the vortex ring. Our analyses of various vortex-ring events all confirm the superior fidelity of the S2W model as compared to the other two models.

\section*{Other intriguing observations}\label{SecIV}
Besides model testing, we have also uncovered other intriguing phenomena in our experiment. For instance, sometimes we see vortex rings that are heavily doped with D$_2$ particles spontaneously flip to the downward direction. A collection of such events are included in Supplementary Video 5. In Fig.~\ref{Fig3}a, we superimpose the images of a representative ring taken at different $t$ to show how the ring changes its direction while it shrinks. This phenomenon can be understood by noting that the vortex ring carries a momentum~\cite{Donnelly-1991-B} $\mathbf{P}(t)=\rho_s\kappa\pi R(t)^2 \mathbf{\hat{n}}$, where $\mathbf{\hat{n}}$ is the unit vector normal to the ring plane pointing in the direction of the ring's motion. The mutual friction and the Stokes drag constantly reduce the ring's momentum, resulting in the shrinkage of the ring. On the other hand, the gravitational force from the trapped particles continuously generates momentum in the downward direction, which forces the ring to flip downward. To test this physical picture, we have conducted simulations using the S2W model. For a heavily doped ring, the exact number $N$ and the radiuses of the trapped particles are hard to determine. Instead, we assume the same radius $a$ for all the trapped particles and treat both $a$ and $N$ as adjustable parameters. For the ring trajectory presented in Fig.~\ref{Fig3}a, we find that it can be reasonably reproduced with $N=36$ and $a=4.9$~$\mu$m, as shown in Fig.~\ref{Fig3}b.

Another intriguing observation is related to the destiny of the particles on the vortex rings. As a ring shrinks, we always see that the trapped particles form a cluster and suddenly switch from the high-speed motion to slowly falling in He II (see Supplementary Video 6). Fig.~\ref{Fig3}c shows an event where the ring plane is nearly perpendicular to the laser plane. Nonetheless, we can measure the centroid velocity $u_p(t)$ of the particles. As shown in Fig.~\ref{Fig3}d, $u_p(t)$ increases drastically as the ring shrinks. At $t=3.375$~s, the trapped particles aggregate to a single cluster and $u_p(t)$ suddenly drops to the expected settling velocity of about 0.1~mm/s. Our interpretation of this phenomenon is that as the ring shrinks, its velocity relative to the normal fluid becomes so large such that the Stokes drag can pull the trapped particles off the vortex core. Subsequently, the bare ring moves away and diminishes, while the left-behind particles form a cluster that decelerates rapidly to the settling velocity due to the Stokes drag. This hypothesis can be tested by comparing the maximum trapping force on a particle from the vortex core (i.e., estimated as~\cite{Meichle-2014-RSI} $f_v\simeq\rho_s\kappa^2/3\pi$) with the Stokes drag $f_D\simeq6\pi a \mu_n u_p$. For our particles with a mean radius $a\simeq1$~$\mu$m, $f_D$ becomes greater than $f_v$ when $u_p$ reaches a threshold value of 5.1~mm/s. This threshold $u_p$ is close to the observed maximum $u_p$ in Fig.~\ref{Fig3}d, which provides a clear support to our understanding. Future systematic studies of the data may provide us deeper insights on the particle-vortex interaction.

\section*{Discussion}\label{SecV}
The results that we have presented provide the first-ever evidence proving that the S2W model can precisely account for the mutual-friction dissipation experienced by quantized vortices in He II. This study may stimulate extensive future research in two directions. First, the S2W model does not rely on empirical experimental inputs and therefore can be readily adapted for other quantum two-fluid systems. An accurate evaluation of the mutual friction is particularly important for processes that involve rapid motion of the quantized vortices, such as vortex reconnections, and pinning and depinning of vortices on solid boundaries. The latter process is the key for understanding glitches in neutron star rotation~\cite{Greenstein-1970-Nature,Packard-1972-PRL}. Our validation of the S2W model therefore paves the way for future high-fidelity simulations of these important processes. The second direction is to examine how the implementation of the S2W model may alter our existing knowledge on quantum turbulence (QT) induced by a chaotic tangle of quantized vortices. For instance, an important topic in QT research is counterflow turbulence where the mutual friction exists at all length scales~\cite{Gao-2017-PRB,Bao-2018-PRB}. Our knowledge on the vortex-tangle properties\cite{Schwarz-1988-PRB,Adachi-2010-PRB}, disturbances in the normal fluid\cite{Mastracci-2019-PRF-2,Yui-2020-PRL}, and the effect of the mutual friction on the mean-velocity profile~\cite{Marakov-2015-PRB,Yui-2018-PRL,Pomyalov-2020-PRB} may subject to change with future S2W simulations.

\section*{References}
\bibliographystyle{naturemag}
\bibliography{Ring}

\begin{thebibliography}{10}
\expandafter\ifx\csname url\endcsname\relax
  \def\url#1{\texttt{#1}}\fi
\expandafter\ifx\csname urlprefix\endcsname\relax\def\urlprefix{URL }\fi
\providecommand{\bibinfo}[2]{#2}
\providecommand{\eprint}[2][]{\url{#2}}

\bibitem{Landau-book}
\bibinfo{author}{Landau, L.~D.} \& \bibinfo{author}{Lifshitz, E.~M.}
\newblock \emph{\bibinfo{title}{Fluid Mechanics}}, vol.~\bibinfo{volume}{6}
  (\bibinfo{publisher}{{Pergamon Press}}, \bibinfo{address}{Oxford},
  \bibinfo{year}{1987}), \bibinfo{edition}{2} edn.

\bibitem{Vinen-1957-PRS-III}
\bibinfo{author}{Vinen, W.~F.}
\newblock \bibinfo{title}{Mutual friction in a heat current in liquid helium
  {II}. {III}. theory of the mutual friction}.
\newblock \emph{\bibinfo{journal}{Proc. Roy. Soc. A}}
  \textbf{\bibinfo{volume}{242}}, \bibinfo{pages}{493--515}
  (\bibinfo{year}{1957}).

\bibitem{Tisza-1938-Nature}
\bibinfo{author}{Tisza, L.}
\newblock \bibinfo{title}{Transport phenomena in helium {II}}.
\newblock \emph{\bibinfo{journal}{Nature}} \textbf{\bibinfo{volume}{141}},
  \bibinfo{pages}{913} (\bibinfo{year}{1938}).

\bibitem{VanSciver-2012-book}
\bibinfo{author}{Van~Sciver, S.~W.}
\newblock \emph{\bibinfo{title}{Helium Cryogenics}}.
\newblock International cryogenics monograph series
  (\bibinfo{publisher}{Springer}, \bibinfo{address}{New York, USA},
  \bibinfo{year}{2012}), \bibinfo{edition}{2} edn.

\bibitem{Schwarz-1977-PRL}
\bibinfo{author}{Schwarz, K.~W.}
\newblock \bibinfo{title}{Theory of turbulence in superfluid
  $^{4}\mathrm{He}$}.
\newblock \emph{\bibinfo{journal}{Phys. Rev. Lett.}}
  \textbf{\bibinfo{volume}{38}}, \bibinfo{pages}{551--554}
  (\bibinfo{year}{1977}).

\bibitem{Schwarz-1988-PRB}
\bibinfo{author}{Schwarz, K.~W.}
\newblock \bibinfo{title}{Three-dimensional vortex dynamics in superfluid
  $^{4}\mathrm{He}$: Homogeneous superfluid turbulence}.
\newblock \emph{\bibinfo{journal}{Phys. Rev. B}} \textbf{\bibinfo{volume}{38}},
  \bibinfo{pages}{2398--2417} (\bibinfo{year}{1988}).

\bibitem{Idowu-2000-PRB}
\bibinfo{author}{Idowu, O.~C.}, \bibinfo{author}{Willis, A.},
  \bibinfo{author}{Barenghi, C.~F.} \& \bibinfo{author}{Samuels, D.~C.}
\newblock \bibinfo{title}{Local normal-fluid helium ii flow due to mutual
  friction interaction with the superfluid}.
\newblock \emph{\bibinfo{journal}{Phys. Rev. B}} \textbf{\bibinfo{volume}{62}},
  \bibinfo{pages}{3409--3415} (\bibinfo{year}{2000}).

\bibitem{Kivotides-2018-PRF}
\bibinfo{author}{Kivotides, D.}
\newblock \bibinfo{title}{Superfluid helium-4 hydrodynamics with discrete
  topological defects}.
\newblock \emph{\bibinfo{journal}{Phys. Rev. Fluids}}
  \textbf{\bibinfo{volume}{3}}, \bibinfo{pages}{104701} (\bibinfo{year}{2018}).

\bibitem{Galantucci-2020-EPJ}
\bibinfo{author}{Galantucci, L.}, \bibinfo{author}{Baggaley, A.~W.},
  \bibinfo{author}{Barenghi, C.~F.} \& \bibinfo{author}{Krstulovic, G.}
\newblock \bibinfo{title}{A new self-consistent approach of quantum turbulence
  in superfluid helium}.
\newblock \emph{\bibinfo{journal}{Eur. Phys. J. Plus}}
  \textbf{\bibinfo{volume}{135}}, \bibinfo{pages}{547} (\bibinfo{year}{2020}).

\bibitem{Yui-2020-PRL}
\bibinfo{author}{Yui, S.}, \bibinfo{author}{Kobayashi, H.},
  \bibinfo{author}{Tsubota, M.} \& \bibinfo{author}{Guo, W.}
\newblock \bibinfo{title}{Fully coupled dynamics of the two fluids in
  superfluid {$^4$He}: Anomalous anisotropic velocity fluctuations in
  counterflow}.
\newblock \emph{\bibinfo{journal}{Phys. Rev. Lett.}}
  \textbf{\bibinfo{volume}{124}}, \bibinfo{pages}{155301}
  (\bibinfo{year}{2020}).

\bibitem{Haljan-2001-PRL}
\bibinfo{author}{Haljan, P.~C.}, \bibinfo{author}{Coddington, I.},
  \bibinfo{author}{Engels, P.} \& \bibinfo{author}{Cornell, E.~A.}
\newblock \bibinfo{title}{Driving {Bose-Einstein-Condensate} vorticity with a
  rotating normal cloud}.
\newblock \emph{\bibinfo{journal}{Phys. Rev. Lett.}}
  \textbf{\bibinfo{volume}{87}}, \bibinfo{pages}{210403}
  (\bibinfo{year}{2001}).

\bibitem{Kwon-2021-Nature}
\bibinfo{author}{Kwon, W.} \emph{et~al.}
\newblock \bibinfo{title}{Sound emission and annihilations in a programmable
  quantum vortex collider}.
\newblock \emph{\bibinfo{journal}{Nature}} \textbf{\bibinfo{volume}{600}},
  \bibinfo{pages}{64--69} (\bibinfo{year}{2021}).

\bibitem{Greenstein-1970-Nature}
\bibinfo{author}{Greenstein, G.}
\newblock \bibinfo{title}{Superfluid turbulence in neutron stars}.
\newblock \emph{\bibinfo{journal}{Nature}} \textbf{\bibinfo{volume}{227}},
  \bibinfo{pages}{791--794} (\bibinfo{year}{1970}).

\bibitem{Packard-1972-PRL}
\bibinfo{author}{Packard, R.~E.}
\newblock \bibinfo{title}{Pulsar speedups related to metastability of the
  superfluid neutron-star core}.
\newblock \emph{\bibinfo{journal}{Phys. Rev. Lett.}}
  \textbf{\bibinfo{volume}{28}}, \bibinfo{pages}{1080--1082}
  (\bibinfo{year}{1972}).

\bibitem{Andersson-2006-MNRAS}
\bibinfo{author}{Andersson, N.}, \bibinfo{author}{Sidery, T.} \&
  \bibinfo{author}{Comer, G.~L.}
\newblock \bibinfo{title}{Mutual friction in superfluid neutron stars}.
\newblock \emph{\bibinfo{journal}{Mon. Not. R. Astron. Soc. Lett.}}
  \textbf{\bibinfo{volume}{368}}, \bibinfo{pages}{162--170}
  (\bibinfo{year}{2006}).

\bibitem{Chesler-2013-Science}
\bibinfo{author}{Chesler, P.~M.}, \bibinfo{author}{Liu, H.} \&
  \bibinfo{author}{Adams, A.}
\newblock \bibinfo{title}{Holographic vortex liquids and superfluid
  turbulence}.
\newblock \emph{\bibinfo{journal}{Science}} \textbf{\bibinfo{volume}{341}},
  \bibinfo{pages}{368--372} (\bibinfo{year}{2013}).

\bibitem{Wittmer-2021-PRL}
\bibinfo{author}{Wittmer, P.}, \bibinfo{author}{Schmied, C.-M.},
  \bibinfo{author}{Gasenzer, T.} \& \bibinfo{author}{Ewerz, C.}
\newblock \bibinfo{title}{Vortex motion quantifies strong dissipation in a
  holographic superfluid}.
\newblock \emph{\bibinfo{journal}{Phys. Rev. Lett.}}
  \textbf{\bibinfo{volume}{127}}, \bibinfo{pages}{101601}
  (\bibinfo{year}{2021}).

\bibitem{Donnelly-1991-B}
\bibinfo{author}{Donnelly, R.~J.}
\newblock \emph{\bibinfo{title}{Quantized vortices in helium II}},
  vol.~\bibinfo{volume}{2} (\bibinfo{publisher}{Cambridge University Press},
  \bibinfo{address}{Cambridge, UK}, \bibinfo{year}{1991}).

\bibitem{Tilley-1990-book}
\bibinfo{author}{Tilley, D.} \& \bibinfo{author}{Tilley, J.}
\newblock \emph{\bibinfo{title}{Superfluidity and Superconductivity}}
  (\bibinfo{publisher}{Institute of Physics}, \bibinfo{address}{Bristol, UK},
  \bibinfo{year}{1990}), \bibinfo{edition}{3} edn.

\bibitem{Vinen-2007-book}
\bibinfo{author}{Vinen, W.~F.} \& \bibinfo{author}{Donnelly, R.~J.}
\newblock \bibinfo{title}{Quantum turbulence}.
\newblock \emph{\bibinfo{journal}{Phys. Today}} \textbf{\bibinfo{volume}{60}},
  \bibinfo{pages}{43--48} (\bibinfo{year}{2007}).

\bibitem{Tsubota-2013-book}
\bibinfo{author}{Tsubota, M.} \& \bibinfo{author}{Kasamatsu, K.}
\newblock \emph{\bibinfo{title}{Quantized Vortices and Quantum Turbulence}},
  \bibinfo{pages}{283--299} (\bibinfo{publisher}{Springer Berlin Heidelberg},
  \bibinfo{address}{Berlin, Germany}, \bibinfo{year}{2013}).

\bibitem{Larbalestier-2001-Nature}
\bibinfo{author}{Larbalestier, D.}, \bibinfo{author}{Gurevich, A.},
  \bibinfo{author}{Feldmann, D.~M.} \& \bibinfo{author}{Polyanskii, A.}
\newblock \bibinfo{title}{High-\textit{T}$_{c}$ superconducting materials for
  electric power applications}.
\newblock \emph{\bibinfo{journal}{Nature}} \textbf{\bibinfo{volume}{414}},
  \bibinfo{pages}{368--377} (\bibinfo{year}{2001}).

\bibitem{Zurek-1985-Nature}
\bibinfo{author}{Zurek, W.~H.}
\newblock \bibinfo{title}{Cosmological experiments in superfluid helium?}
\newblock \emph{\bibinfo{journal}{Nature}} \textbf{\bibinfo{volume}{317}},
  \bibinfo{pages}{505--508} (\bibinfo{year}{1985}).

\bibitem{Adachi-2010-PRB}
\bibinfo{author}{Adachi, H.}, \bibinfo{author}{Fujiyama, S.} \&
  \bibinfo{author}{Tsubota, M.}
\newblock \bibinfo{title}{Steady-state counterflow quantum turbulence:
  Simulation of vortex filaments using the full biot-savart law}.
\newblock \emph{\bibinfo{journal}{Phys. Rev. B}} \textbf{\bibinfo{volume}{81}},
  \bibinfo{pages}{104511} (\bibinfo{year}{2010}).

\bibitem{Risto-2014-PNAS}
\bibinfo{author}{H{\"a}nninen, R.} \& \bibinfo{author}{Baggaley, A.~W.}
\newblock \bibinfo{title}{Vortex filament method as a tool for computational
  visualization of quantum turbulence}.
\newblock \emph{\bibinfo{journal}{Proc. Natl. Acad. Sci. U.S.A.}}
  \textbf{\bibinfo{volume}{111}}, \bibinfo{pages}{4667--4674}
  (\bibinfo{year}{2014}).

\bibitem{Yui-2022-PRL}
\bibinfo{author}{Yui, S.}, \bibinfo{author}{Tang, Y.}, \bibinfo{author}{Guo,
  W.}, \bibinfo{author}{Kobayashi, H.} \& \bibinfo{author}{Tsubota, M.}
\newblock \bibinfo{title}{Universal anomalous diffusion of quantized vortices
  in ultraquantum turbulence}.
\newblock \emph{\bibinfo{journal}{Phys. Rev. Lett.}}
  \textbf{\bibinfo{volume}{129}}, \bibinfo{pages}{025301}
  (\bibinfo{year}{2022}).

\bibitem{Yui-2018-PRL}
\bibinfo{author}{Yui, S.}, \bibinfo{author}{Tsubota, M.} \&
  \bibinfo{author}{Kobayashi, H.}
\newblock \bibinfo{title}{Three-dimensional coupled dynamics of the two-fluid
  model in superfluid $^{4}\mathrm{He}$: Deformed velocity profile of normal
  fluid in thermal counterflow}.
\newblock \emph{\bibinfo{journal}{Phys. Rev. Lett.}}
  \textbf{\bibinfo{volume}{120}}, \bibinfo{pages}{155301}
  (\bibinfo{year}{2018}).

\bibitem{Kivotides-2000-Science}
\bibinfo{author}{Kivotides, D.}, \bibinfo{author}{Barenghi, C.~F.} \&
  \bibinfo{author}{Samuels, D.~C.}
\newblock \bibinfo{title}{Triple vortex ring structure in superfluid helium
  {II}}.
\newblock \emph{\bibinfo{journal}{Science}} \textbf{\bibinfo{volume}{290}},
  \bibinfo{pages}{777--779} (\bibinfo{year}{2000}).

\bibitem{Bewley-2006-Nature}
\bibinfo{author}{Bewley, G.~P.}, \bibinfo{author}{Lathrop, D.~P.} \&
  \bibinfo{author}{Sreenivasan, K.~R.}
\newblock \bibinfo{title}{Superfluid helium: Visualization of quantized
  vortices}.
\newblock \emph{\bibinfo{journal}{Nature}} \textbf{\bibinfo{volume}{441}},
  \bibinfo{pages}{588} (\bibinfo{year}{2006}).

\bibitem{Guo-2014-PNAS}
\bibinfo{author}{Guo, W.}, \bibinfo{author}{La~Mantia, M.},
  \bibinfo{author}{Lathrop, D.~P.} \& \bibinfo{author}{Van~Sciver, S.~W.}
\newblock \bibinfo{title}{Visualization of two-fluid flows of superfluid
  helium-4}.
\newblock \emph{\bibinfo{journal}{Proc. Natl. Acad. Sci. U.S.A}}
  \textbf{\bibinfo{volume}{111}}, \bibinfo{pages}{4653--4658}
  (\bibinfo{year}{2014}).

\bibitem{La_Mantia-2014-EL}
\bibinfo{author}{Mantia, M.~L.} \& \bibinfo{author}{Skrbek, L.}
\newblock \bibinfo{title}{Quantum, or classical turbulence?}
\newblock \emph{\bibinfo{journal}{Europhys. Lett.}}
  \textbf{\bibinfo{volume}{105}}, \bibinfo{pages}{46002}
  (\bibinfo{year}{2014}).

\bibitem{Mastracci-2019-PRF}
\bibinfo{author}{Mastracci, B.} \& \bibinfo{author}{Guo, W.}
\newblock \bibinfo{title}{Characterizing vortex tangle properties in
  steady-state {He II} counterflow using particle tracking velocimetry}.
\newblock \emph{\bibinfo{journal}{Phys. Rev. Fluids}}
  \textbf{\bibinfo{volume}{4}}, \bibinfo{pages}{023301} (\bibinfo{year}{2019}).

\bibitem{Fonda-2019-PNAS}
\bibinfo{author}{Fonda, E.}, \bibinfo{author}{Sreenivasan, K.~R.} \&
  \bibinfo{author}{Lathrop, D.~P.}
\newblock \bibinfo{title}{Reconnection scaling in quantum fluids}.
\newblock \emph{\bibinfo{journal}{Proc. Natl. Acad. Sci. U.S.A}}
  \textbf{\bibinfo{volume}{116}}, \bibinfo{pages}{1924--1928}
  (\bibinfo{year}{2019}).

\bibitem{Tang-2021-PNAS}
\bibinfo{author}{Tang, Y.}, \bibinfo{author}{Bao, S.} \& \bibinfo{author}{Guo,
  W.}
\newblock \bibinfo{title}{Superdiffusion of quantized vortices uncovering
  scaling laws in quantum turbulence}.
\newblock \emph{\bibinfo{journal}{Proc. Natl. Acad. Sci. U.S.A}}
  \textbf{\bibinfo{volume}{118}}, \bibinfo{pages}{e2021957118}
  (\bibinfo{year}{2021}).

\bibitem{Bewley-2009-JLTP}
\bibinfo{author}{Bewley, G.~P.} \& \bibinfo{author}{Sreenivasan, K.~R.}
\newblock \bibinfo{title}{The decay of a quantized vortex ring and the
  influence of tracer particles}.
\newblock \emph{\bibinfo{journal}{J. Low Temp. Phys.}}
  \textbf{\bibinfo{volume}{156}}, \bibinfo{pages}{84--94}
  (\bibinfo{year}{2009}).

\bibitem{Mastracci-2018-RSI}
\bibinfo{author}{Mastracci, B.} \& \bibinfo{author}{Guo, W.}
\newblock \bibinfo{title}{An apparatus for generation and quantitative
  measurement of homogeneous isotropic turbulence in {He II}}.
\newblock \emph{\bibinfo{journal}{Rev. Sci. Instrum.}}
  \textbf{\bibinfo{volume}{89}}, \bibinfo{pages}{015107}
  (\bibinfo{year}{2018}).

\bibitem{Tang-2020-PRF}
\bibinfo{author}{Tang, Y.}, \bibinfo{author}{Bao, S.}, \bibinfo{author}{Kanai,
  T.} \& \bibinfo{author}{Guo, W.}
\newblock \bibinfo{title}{Statistical properties of homogeneous and isotropic
  turbulence in {He} {II} measured via particle tracking velocimetry}.
\newblock \emph{\bibinfo{journal}{Phys. Rev. Fluids}}
  \textbf{\bibinfo{volume}{5}}, \bibinfo{pages}{084602} (\bibinfo{year}{2020}).

\bibitem{Sbalzarini-2005-JSB}
\bibinfo{author}{Sbalzarini, I.~F.} \& \bibinfo{author}{Koumoutsakos, P.}
\newblock \bibinfo{title}{Feature point tracking and trajectory analysis for
  video imaging in cell biology}.
\newblock \emph{\bibinfo{journal}{J. Struct. Biol}}
  \textbf{\bibinfo{volume}{151}}, \bibinfo{pages}{182--195}
  (\bibinfo{year}{2005}).

\bibitem{Skoblin-2020-JLTP}
\bibinfo{author}{Skoblin, A.~A.}, \bibinfo{author}{Zlenko, D.~V.} \&
  \bibinfo{author}{Stovbun, S.~V.}
\newblock \bibinfo{title}{Friction force limits the drift of microparticles
  along the quantum vortex in liquid helium}.
\newblock \emph{\bibinfo{journal}{J. Low Temp. Phys.}}
  \textbf{\bibinfo{volume}{200}}, \bibinfo{pages}{91--101}
  (\bibinfo{year}{2020}).

\bibitem{Mineda-2013-PRB}
\bibinfo{author}{Mineda, Y.}, \bibinfo{author}{Tsubota, M.},
  \bibinfo{author}{Sergeev, Y.~A.}, \bibinfo{author}{Barenghi, C.~F.} \&
  \bibinfo{author}{Vinen, W.~F.}
\newblock \bibinfo{title}{Velocity distributions of tracer particles in thermal
  counterflow in superfluid ${}^{4}$he}.
\newblock \emph{\bibinfo{journal}{Phys. Rev. B}} \textbf{\bibinfo{volume}{87}},
  \bibinfo{pages}{174508} (\bibinfo{year}{2013}).

\bibitem{Barenghi-2009-PRB}
\bibinfo{author}{Barenghi, C.~F.} \& \bibinfo{author}{Sergeev, Y.~A.}
\newblock \bibinfo{title}{Motion of vortex ring with tracer particles in
  superfluid helium}.
\newblock \emph{\bibinfo{journal}{Phys. Rev. B}} \textbf{\bibinfo{volume}{80}},
  \bibinfo{pages}{024514} (\bibinfo{year}{2009}).

\bibitem{Meichle-2014-RSI}
\bibinfo{author}{Meichle, D.~P.} \& \bibinfo{author}{Lathrop, D.~P.}
\newblock \bibinfo{title}{Nanoparticle dispersion in superfluid helium}.
\newblock \emph{\bibinfo{journal}{Rev. Sci. Instrum.}}
  \textbf{\bibinfo{volume}{85}}, \bibinfo{pages}{073705}
  (\bibinfo{year}{2014}).

\bibitem{Gao-2017-PRB}
\bibinfo{author}{Gao, J.}, \bibinfo{author}{Varga, E.}, \bibinfo{author}{Guo,
  W.} \& \bibinfo{author}{Vinen, W.~F.}
\newblock \bibinfo{title}{Energy spectrum of thermal counterflow turbulence in
  superfluid helium-4}.
\newblock \emph{\bibinfo{journal}{Phys. Rev. B}} \textbf{\bibinfo{volume}{96}},
  \bibinfo{pages}{094511} (\bibinfo{year}{2017}).

\bibitem{Bao-2018-PRB}
\bibinfo{author}{Bao, S.}, \bibinfo{author}{Guo, W.}, \bibinfo{author}{L'vov,
  V.~S.} \& \bibinfo{author}{Pomyalov, A.}
\newblock \bibinfo{title}{Statistics of turbulence and intermittency
  enhancement in superfluid $^{4}\mathrm{He}$ counterflow}.
\newblock \emph{\bibinfo{journal}{Phys. Rev. B}} \textbf{\bibinfo{volume}{98}},
  \bibinfo{pages}{174509} (\bibinfo{year}{2018}).

\bibitem{Mastracci-2019-PRF-2}
\bibinfo{author}{Mastracci, B.}, \bibinfo{author}{Bao, S.},
  \bibinfo{author}{Guo, W.} \& \bibinfo{author}{Vinen, W.~F.}
\newblock \bibinfo{title}{Particle tracking velocimetry applied to thermal
  counterflow in superfluid $^{4}\mathrm{He}$: Motion of the normal fluid at
  small heat fluxes}.
\newblock \emph{\bibinfo{journal}{Phys. Rev. Fluids}}
  \textbf{\bibinfo{volume}{4}}, \bibinfo{pages}{083305} (\bibinfo{year}{2019}).

\bibitem{Marakov-2015-PRB}
\bibinfo{author}{Marakov, A.} \emph{et~al.}
\newblock \bibinfo{title}{Visualization of the normal-fluid turbulence in
  counterflowing superfluid $^{4}${{He}}}.
\newblock \emph{\bibinfo{journal}{Phys. Rev. B}} \textbf{\bibinfo{volume}{91}},
  \bibinfo{pages}{094503} (\bibinfo{year}{2015}).

\bibitem{Pomyalov-2020-PRB}
\bibinfo{author}{Pomyalov, A.}
\newblock \bibinfo{title}{Dynamics of turbulent plugs in a superfluid
  $^{4}\mathrm{He}$ channel counterflow}.
\newblock \emph{\bibinfo{journal}{Phys. Rev. B}}
  \textbf{\bibinfo{volume}{101}}, \bibinfo{pages}{134515}
  (\bibinfo{year}{2020}).

\bibitem{Press-1992-book}
\bibinfo{author}{Press, W.~H.}, \bibinfo{author}{Flannery, B.~P.},
  \bibinfo{author}{Teukolsky, S.~A.} \& \bibinfo{author}{Vetterling, W.~T.}
\newblock \emph{\bibinfo{title}{Numerical Recipes in C. The Art of Scientific
  Computing}} (\bibinfo{publisher}{Cambridge University Press},
  \bibinfo{address}{Cambridge}, \bibinfo{year}{1992}).

\end{thebibliography}

\section*{Methods}\label{SecVI}
\noindent\textbf{Numerical models}\\
\noindent\emph{Schwarz model}: In the framework of Schwarz's vortex filament model~\cite{Schwarz-1988-PRB}, all the quantized vortex lines are represented by zero-thickness filaments. The position vector of a filament can be written in the parametric form $\mathbf{s}=\mathbf{s}(\xi,t)$, where $\xi$ denotes the arc length along the filament. In the presence of the viscous normal fluid, a short segment $\Delta\xi$ of a vortex filament located at $\mathbf{s}$ would experience two forces, i.e., the Magnus force $\mathbf{f}_M=\rho_s\kappa \mathbf{s}'\times(\mathbf{u}_L-\mathbf{u}_s)\Delta \xi$ and the mutual friction force $\mathbf{f}_{sn}=[-\gamma_0\mathbf{s}'\times(\mathbf{s}'\times(\mathbf{u}_n-\mathbf{u}_L))+\gamma_0'\mathbf{s}'\times(\mathbf{u}_n-\mathbf{u}_L)]\Delta \xi$. By balancing these two forces, the velocity of this segment $\mathbf{u}_L=d\mathbf{s}/dt$ can be derived as:
\begin{equation}
d\mathbf{s}/dt=\mathbf{u}_s+\alpha\mathbf{s}'\times(\mathbf{u}_n-\mathbf{u}_s)-\alpha'\mathbf{s}'\times[\mathbf{s}'\times(\mathbf{u}_n-\mathbf{u}_s)],
\label{Schwarz}
\end{equation}
where the coefficients $\alpha$ and $\alpha'$ depend on the empirical mutual friction coefficients $\gamma_0$ and $\gamma_0'$, whose values have been tabulated~\cite{Donnelly-1991-B}. While the normal-fluid velocity $\mathbf{u}_n$ is prescribed, the local superfluid velocity $\mathbf{u}_s(\mathbf{s},t)$ is evaluated as the sum of the background flow velocity $\mathbf{u}_{s0}$ and the velocity $\mathbf{u}_{in}$ induced at $\mathbf{s}$ by all the vortices, which can be calculated using the full Boit-Savart integral~\cite{Adachi-2010-PRB}:
\begin{equation}
\mathbf{u}_{in}(\mathbf{s},t)=\frac{\kappa}{4\pi}\int\frac{(\mathbf{s_1}-\mathbf{s})\times d\mathbf{s_1}}{|\mathbf{s_1}-\mathbf{s}|^3},
\end{equation}
where the integration goes over all the vortex filaments. When we apply the Schwarz model to simulate the motion of a vortex ring in quiescent He II, we set both $\mathbf{u}_n$ and $\mathbf{u}_{s0}$ to zero and discretize the initial ring with a resolution $\Delta\xi=0.005$~mm. The time evolution of each vortex segment's position can then be obtained through a temporal integration of Eq.~(\ref{Schwarz}) using the fourth-order Runge-Kutta method~\cite{Press-1992-book} with a time step $\Delta t=10^{-5}$ s.

\emph{2W model}: In the 2W model, the normal-fluid velocity $\mathbf{u}_n$ is no longer prescribed. Instead, it is calculated by solving the classical Navier-Stokes equation with an added mutual friction term~\cite{Yui-2020-PRL}:
\begin{equation}
\frac{\partial \mathbf{u}_n}{\partial t}+(\mathbf{u}_n\cdot\mathbf{\nabla})\mathbf{u}_n=-\frac{1}{\rho_{\mathrm{He}}}\mathbf{\nabla}P+\nu_n\nabla^2\mathbf{u}_n+\frac{\mathbf{F}_{ns}}{\rho_n}
\label{2W}
\end{equation}
where $\rho_n$ and $\rho_{\mathrm{He}}$ are, respectively, the normal-fluid density and the total density of He II, $P$ is the pressure, $\nu_n$ is the He II kinematic viscosity, and $\mathbf{F}_{ns}$ is the mutual friction per unit volume which can be calculated as:
\begin{equation}
\mathbf{F}_{ns}(\mathbf{r})=\frac{1}{\Delta\Omega(\mathbf{r})}\int_\mathcal{L(\mathbf{r})}(-\mathbf{f}_{sn}/\Delta\xi)d\xi
\end{equation}
where $\mathcal{L(\mathbf{r})}$ denotes that the integration is performed along all the vortex lines in the computational cell $\Delta\Omega(\mathbf{r})=\Delta x\times\Delta y\times\Delta z$ located at $\mathbf{r}$. When we simulate the vortex ring dynamics, Eq.~(\ref{Schwarz}) and Eq.~(\ref{2W}) are solved together to render the positions of the vortex-ring segments $\mathbf{s}(\xi,t)$ and the normal-fluid velocity $\mathbf{u}_n$. The time integration of Eq.~(\ref{2W}) is conducted using the second-order Adams-Bashforth method~\cite{Yui-2020-PRL} with the same time step $\Delta t$, and the spatial differentiation is performed via the second-order finite difference with a spatial resolution $\Delta x=\Delta y=\Delta z$=0.008~mm.

\emph{S2W model}: In the S2W model, the mutual friction force that acts on a vortex segment $\Delta\xi$ is given by~\cite{Galantucci-2020-EPJ}:
\begin{equation}
\mathbf{f}_{sn}=[-D\mathbf{s}'\times(\mathbf{s}'\times(\mathbf{u}_n-\mathbf{u}_L))-\rho_n\kappa\mathbf{s}'\times(\mathbf{u}_n-\mathbf{u}_L)]\Delta\xi,
\end{equation}
where the only friction coefficient $D$ can be calculated as:
\begin{equation}
D=-4\pi\rho_n\nu_n/[0.0772+\ln(|\mathbf{u}^{\perp}_n-\mathbf{u}_L|a_0/4\nu_n)].
\label{D}
\end{equation}
Here $a_0\simeq1$~{\AA} is the vortex-core radius and $\mathbf{u}^{\perp}_n$ denotes the local normal-fluid velocity at the vortex-segment location that is projected in the plane perpendicular to the segment~\cite{Galantucci-2020-EPJ}. By balancing the Magnus force $\mathbf{f}_M$ and the revised mutual friction force, the equation of motion for the vortex segment is now given by:
\begin{equation}
d\mathbf{s}/dt=\mathbf{u}_s+\beta\mathbf{s}'\times(\mathbf{u}_n-\mathbf{u}_s)-\beta'\mathbf{s}'\times[\mathbf{s}'\times(\mathbf{u}_n-\mathbf{u}_s)],
\label{S2W}
\end{equation}
where the coefficients $\beta$ and $\beta'$ depends on $D$ as derived by Galantucci \emph{et al}.~\cite{Galantucci-2020-EPJ}. The evolution of the vortex position and $\mathbf{u}_n$ can be obtained by solving Eq.~(\ref{2W}) and Eq.~(\ref{S2W}) with $D$ evaluated self-consistently via Eq.~(\ref{D}).

For a quantized vortex ring with a radius $R$ moving in quiescent He II, the self-induced superfluid velocity at the ring's location is given by~\cite{Donnelly-1991-B} $\mathbf{u}_s=\frac{\kappa}{4\pi R}[\ln(8R/a_0)-\frac{1}{2}]\mathbf{\hat{n}}$, which is the same in all three models. However, the local $\mathbf{u}_n$ is different, which leads to the different mutual friction dissipation rate. In the Schwarz model, $\mathbf{u}_n=0$ and therefore the highest mutual friction dissipation is expected. In both the 2W model and the S2W model, the back action of the mutual friction in the normal fluid generates two oppositely polarized normal-fluid vortex rings as shown in Fig.~\ref{Fig1}. In the 2W model, the two normal-fluid rings are concentrically located nearly in the same plane as the quantized vortex ring, whereas in the S2W model the two normal-fluid rings are slightly shifted to above and below the quantized-ring plane. This shift changes the direction of the local $\mathbf{u}_n$. Nonetheless, the induced local $\mathbf{u}_n$ in both models has a component in the same direction as the local $\mathbf{u}_s$, which effectively reduces the mutual friction dissipation as compared to that in the Schwarz model.\\

\noindent\textbf{Effects of the trapped particles}\\
\noindent When a vortex segment $\Delta \xi$ carries a trapped particle with a radius $a$, its equation of motion changes to~\cite{Mineda-2013-PRB}:
\begin{equation}
(m_p+m_f)\frac{d\mathbf{u}_L}{dt}=\mathbf{f}_M+\mathbf{f}_{sn}+\mathbf{f}_{D}+\mathbf{f}_{g}
\label{Particle}
\end{equation}
where the term on the left-hand side represents the inertial effect caused by the trapped particle's mass $m_p=\rho_p\frac{4}{3}\pi a^3$ and the fluid's added mass $m_f=\frac{1}{2}\rho_{\mathrm{He}}\frac{4}{3}\pi a^3$. On the right-hand side, besides the Magnus force $\mathbf{f}_M$ and the mutual friction force $\mathbf{f}_{sn}$, two additional forces are included, i.e., the Stokes drag exerted by the normal fluid on the particle $\mathbf{f}_{D}=-6\pi a\mu_n(\mathbf{u}_L-\mathbf{u}_n)$ and the gravitational force $\mathbf{f}_{g}=(\rho_p-\rho_{\mathrm{He}})\frac{4}{3}\pi a^3$. Other minor effects associated with the acceleration of the superfluid and the normal fluid around the trapped particle are negligible~\cite{Mineda-2013-PRB}. This model is accurate when $a$ is much smaller than the separation distance between the particles trapped along the vortex ring, which holds true for the ring events that we selected to analyze.

To get a sense on how large the particle affects are, one may compare the total Stokes drag $F_{D}=|\sum_i\mathbf{f}_{D,i}|$ and the total gravitational force $F_{g}=|\sum_i\mathbf{f}_{g,i}|$ with the total mutual friction force $F_{sn}=|\oint(\mathbf{f}_{sn}/\Delta\xi)d\xi|$, where $\sum_i$ means the summation over all the trapped particles and $\oint$ denotes the integration along the vortex ring. For the 9-particle vortex-ring event shown in Fig.~\ref{Fig1}, using the particle radiuses obtained through the size analysis (see later discussions in Methods), we estimate that $F_{D}/F_{sn}\simeq10\%$ and $F_{g}/F_{sn}\simeq4\%$ at $R(0)=312$~$\mu$m. On the other hand, for the vortex ring shown in Fig.~\ref{Fig2} that carries 2 particles, the estimated ratios are only $F_{D}/F_{sn}\simeq4.8\%$ and $F_{g}/F_{sn}\simeq0.8\%$, despite the ring's smaller initial radius (i.e., $R(0)=140.8$~$\mu$m) and hence higher propagation speed.\\


\renewcommand{\figurename}{\textbf{Extended Data Fig.}}
\setcounter{figure}{0}
\begin{figure}[t]
\centering
\includegraphics[width=1\linewidth]{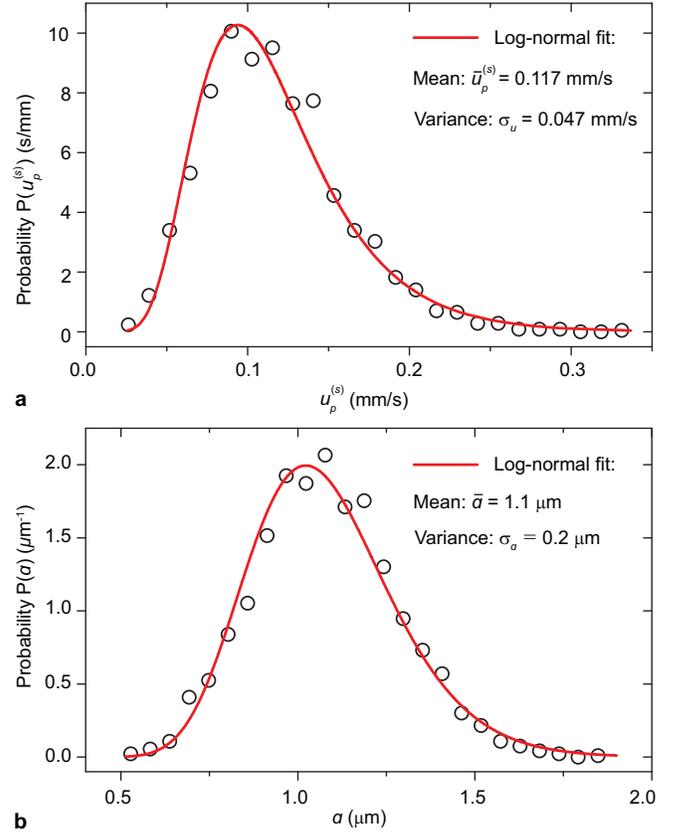}
\caption{\textbf{Settling velocity and radius distribution of the D$_2$ particles in He II at \emph{T}=1.65~K.} \textbf{a}, Distribution of the settling velocity $u^{(s)}_p$. \textbf{b}, Distribution of the particle radius $a$.}
\label{Ext1}
\end{figure}

\noindent\textbf{Particle size distribution}\\
\noindent We produce solidified D$_2$ tracer particles in He II by slowly injecting a mixture of 5\% D$_2$ gas and 95\% $^4$He gas directly into the plexiglass channel immersed in the He II bath. A computer-controlled solenoid valve is used to adjust the injection duration, and a needle valve is adopted to restrict the gas flow rate. Upon the injection, the D$_2$ gas forms solidified ice particles. To evaluate the sizes of the resulted particles, we took images of the particles undergoing freely settling in quiescent He II (see Supplementary Video 7). By tracking the particles in such videos, we can generate a probability distribution of the particle settling velocity $u^{(s)}_p$. The result for $T=1.65$~K is shown in the Extended Data Fig.~\ref{Ext1}a. The $u^{(s)}_p$ data can be fitted nicely with a log-normal distribution, from which we can determine that the distribution is peaked at about 0.1~mm/s.

Note that the settling velocity is achieved when the Stokes drag exerting on a D$_2$ particle is balanced by the gravitational force, i.e., $6\pi a\mu_n u^{(s)}_p=\frac{4\pi}{3}a^3(\rho_p-\rho_{\mathrm{He}})g$. This balance leads to $a=[9\mu_nu^{(s)}_p/2(\rho_p-\rho_{\mathrm{He}})g]^{1/2}$. Therefore, knowing the distribution of $u^{(s)}_p$, we can then generate the radius distribution of the D$_2$ particles. As shown in Extended Data Fig.~\ref{Ext1}b, this distribution is peaked at $a\simeq1.1$~$\mu$m with a variance of about 0.2~$\mu$m.\\

\renewcommand{\tablename}{\textbf{Extended Data Table.}}
\begin{table}[t]	
	\centering
	\caption{Radiuses and initial positions of the trapped particles for the 9-particle vortex ring in Fig.~\ref{Fig1} and the 2-particle vortex ring in Fig.~\ref{Fig2}.}
	\label{scale}
	\begin{center}
		\setlength{\tabcolsep}{2mm}{
		\begin{tabular}{l l l l l}		
		\hline\hline
		9-p ring           &   $x$~(mm)    &   $y$~(mm) &   $z$~(mm) &   $a$~($\mu$m)   \\[5pt]
		\hline
		P1                &   -0.27       &   -0.22    &  0.20      &     0.87          \\[5pt]
        P2                &   -0.19       &   0.25     &  -0.07      &     1.32          \\[5pt]
		P3                &   -0.10       &   -0.28    &  0.22      &     1.04          \\[5pt]
		P4                &   -0.09        &   0.26     &  -0.09      &     1.69          \\[5pt]
		P5                &   -0.01        &   0.24     &  -0.09      &     1.74          \\[5pt]
		P6                &   0.15        &   0.14     &  -0.05      &     1.03          \\[5pt]
        P7                &   0.22        &   0.03     &  0.02      &     0.92          \\[5pt]
		P8                &   -0.36       &   0.15     &  0.01      &     0.78          \\[5pt]
		P9                &   0.22        &   -0.06    &  0.06      &     1.09          \\[5pt]
        \hline\hline
        2-p ring           &   $x$~(mm)    &   $y$~(mm) &   $z$~(mm) &   $a$~($\mu$m)   \\[5pt]
		\hline
		P1                &     -0.01      &   --    &    0.05    &     1.18          \\[5pt]
        P2                &     -0.27      &   --    &    -0.05    &     1.12          \\[5pt]		
		\hline\hline
		\end{tabular}
        }
	\end{center}	
\label{Tab1}
\end{table}

\renewcommand{\figurename}{\textbf{Extended Data Fig.}}
\begin{figure}[t]
\centering
\includegraphics[width=1\linewidth]{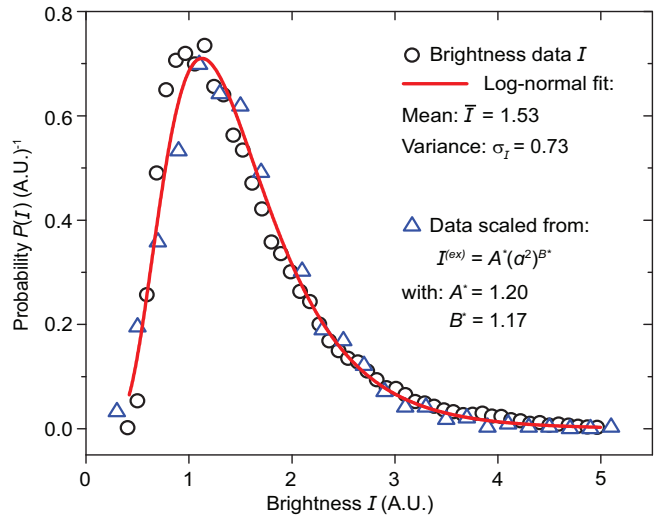}
\caption{\textbf{Distribution of the brightness $I$ of the $D_2$ particles.} The black circles represent the measured brightness $I$. The blue triangles are $I^{(ex)}$ calculated using the distribution of $a$, where $A^*=1.20$ and $B^*=1.17$ are the optimal correlation parameters that render the best agreement between the two distributions.}
\label{Ext2}
\end{figure}

\noindent\textbf{Positions and radiuses of trapped particles}\\
\noindent To evaluate the effects of the trapped $D_2$ particles on the motion of a vortex ring, we need to know the radius and initial position of each individual trapped particle. Using the feature-point tracking routine~\cite{Sbalzarini-2005-JSB}, we can determine the coordinates of every particles in the $x$-$z$ image plane. For particles trapped on the vortex ring, their coordinates ($x_i$,$z_i$) should satisfy the following equation of an ellipse:
\begin{equation}
\begin{split}
&\frac{[(x_i-x_0)\cos\phi+(z_i-z_0)\sin\phi]^2}{R_1^2}\\
&+\frac{[(z_i-z_0)\cos\phi-(x_i-x_0)\sin\phi]^2}{R_2^2}=1,
\label{Ellipse}
\end{split}
\end{equation}
where ($x_0$,$z_0$) are the coordinates of the ellipse center, $R_1$ and $R_2$ are, respectively, the semi-major and semi-minor axes of the ellipse, and $\phi$ is the angle between the ellipse major axis and the $x$-axis. These five parameters can be uniquely determined through a least squares fit to the positions of the trapped particles when there are at least five particles on the ring. Through this fit, we can determine the vortex ring radius $R=R_1$ and the projection angle $\theta$ between the ring's normal vector $\mathbf{\hat{n}}$ and the $x$-$z$ plane (i.e., $\sin\theta=R_2/R_1$). If we set $y_0=0$ for the ellipse center at $t=0$, the initial $y_i$ of each trapped particle can be calculated as $y_i=[(x_i-x_0)\sin\phi-(z_i-z_0)\cos\phi]/\tan\theta$. In the Extended Data Table~\ref{Tab1}, we list the 3D coordinates of all the nine trapped particles for the vortex ring presented in Fig.~\ref{Fig1}. These coordinates are used in our model simulations.

To evaluate the trapped particle's radius $a$, we develop a correlation between $a$ and the particle's brightness $I$. For the particles that undergo freely settling (Supplementary Video 7), we can calculate the brightness $I$ of each particle by summing up the counts in the image pixels associated with the particle. A distribution of the particle brightness $P(I)$ can therefore be generated, which is shown in Extended Data Fig.~\ref{Ext2}. Since $I$ depends on the particle's surface area and hence $a^2$, we can construct a simple correlation $I=A(a^2)^B$, where $A$ and $B$ are tuning parameters. For a given pair $A$ and $B$, we can scale the distribution of $a$ shown in the Extended Data Fig.~\ref{Ext1}b to generate the distribution of the expected brightness $I^{(ex)}=A(a^2)^B$. We then vary $A$ and $B$ to minimize the difference between the $I^{(ex)}$ distribution and the actual distribution $P(I)$. At the optimal values $A^*=1.20$ and $B^*=1.17$, the generated $I^{(ex)}$ distribution agrees nicely with $P(I)$, as shown in Fig.~\ref{Ext2}.

\renewcommand{\figurename}{\textbf{Extended Data Fig.}}
\begin{figure*}[t]
\centering
\includegraphics[width=1\linewidth]{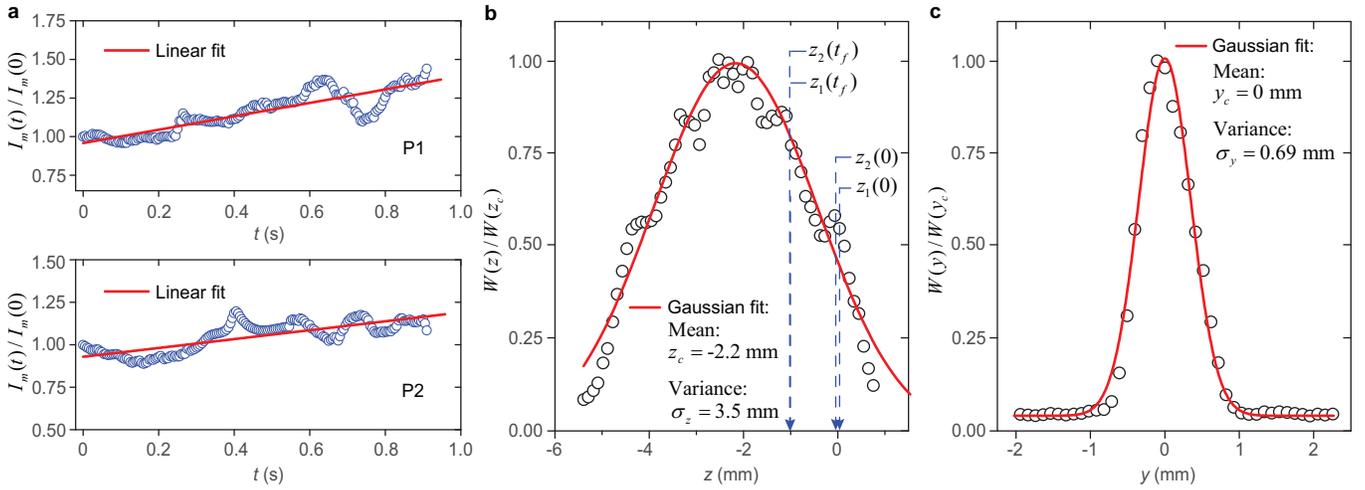}
\caption{\textbf{Trapped particle's brightness variation and laser-intensity cross-sectional profile.} \textbf{a}, Time variation of the directly measured brightness $I_m(t)$ of the two trapped particles as shown in Fig.~\ref{Fig2}a. \textbf{b}, Measured laser intensity $W$ as a function of $z$ (i.e., height direction). The red curve is a Gaussian fit to the data. The $z$-coordinates of the two particles at $t=0$ and $t=t_f$ are indicated. \textbf{c}, Measured laser intensity $W$ as a function of $y$ (i.e., thickness direction).}
\label{Ext3}
\end{figure*}

Using the derived correlation $I=A^*(a^2)^{B^*}$, we can calculate the radius $a_i$ of a trapped particle $i$ by measuring its brightness $I_i$. However, we must note that this correlation holds only in a statistical sense. When we apply it to analyze the radiuses of individual particles, there can be intrinsic uncertainties. For instance, two identical particles can render different brightness (and hence different radiuses) when they are at different locations in the thickness direction of the laser sheet. To improve the reliability, in practice we collect the brightness data of the particle $i$ over the time period that it is observed and then use the time-averaged brightness $\overline{I_i}$ in the correlation to calculate $a_i$. More accurate simulation of the vortex ring's motion can be achieved for rings carrying less amount of trapped particles, such as our 2-particle ring events.\\

\noindent\textbf{Constraint on the projection angle $\theta$}\\
\noindent For the 2-particle vortex ring event presented in Fig.~\ref{Fig2}, a constraint on the projection angle $\theta$ between the ring's propagation direction and the $x$-$z$ image plane can be placed based on the time-variation of the particle's brightness $I(t)$. This is because $\tan\theta=\Delta y/\Delta S$, where $\Delta S=1.12$~mm is the distance traversed by the centroid of the two particles in the $x$-$z$ plane over the observation time $t_f=0.89$~s, and $\Delta y$ is the centroid displacement in the $y$ direction perpendicular to the laser sheet, which can be estimated based on the variation of $I(t)$.

To estimate $\Delta y$, we first show the measured brightness $I_m(t)$ of each particle in the Extended Data Fig.~\ref{Ext3}a. The variation of $I_m(t)$ is caused by the displacement of the particles in both the $y$ direction and the $z$ direction, since $I_m(t)$ is proportional to the laser intensity $W$ which varies primarily in these two directions. To quantify the laser-intensity variations, we then place an optical power meter behind a mask with a narrow slit (20~$\mu$m in width) oriented either horizontally or vertically. By moving the horizontal slit in the $z$ direction or by moving the vertical slit in the $y$ direction, we can measure $W$ as a function of $y$ and $z$. The results are shown in the Extended Data Fig.~\ref{Ext3}b and c, respectively. The profile of $W$ in each direction can be reasonably fit with a Gaussian function, which renders $W(y,z)\propto e^{-2(y-y_c)^2/\sigma^2_y}\cdot e^{-2(z-z_c)^2/\sigma^2_z}$, where $y_c=0$ and $z_c=-2.2$~mm are the coordinates of the beam's cross-sectional center, $\sigma_y=0.69$~mm is the half-thickness of the laser sheet at $1/e^2$ intensity (i.e., which corresponds to a full thickness at half maximum intensity of 0.82~mm), and $\sigma_z=3.5$~mm is the sheet's half-height at $1/e^2$ intensity.

Finally, we can calculate the corrected brightness $I(t)=I_m(t)/e^{-2(z(t)-z_c)^2/\sigma^2_z}$. The results are shown in Fig.~\ref{Fig2}c. The variation of $I(t)$ is entirely due to the particle displacement in the $y$ direction. Since $I(t)/I(0)$ for either particle decreases roughly monotonically by about 20\% over the observation time, we can estimate the displacement $\Delta y$ based on the Extended Data Fig.~\ref{Ext3}c. For a given initial particle coordinate $y(0)$, we can determine $\Delta y$ that gives 20\% laser-intensity drop. By varying $y(0)$, we find that $\Delta y$ can reach up to about 0.2~mm. This sets an upper limit $\tan\theta\leq0.2/1.12=0.18$. Since $\cos\theta=u_p(0)/u(0)$, a constraint on $u(0)$ and hence the initial ring radius $R(0)$ can be placed. This constraint together with the other constraints discussed in the paper render the variation range of the simulated curves as shown in Fig.~\ref{Fig2}d.

\section*{Data Availability}
\noindent The data that support the findings of this study are available from the corresponding author upon reasonable request.

\section*{Code availability}
\noindent All computer codes used in this study are available from the corresponding author upon reasonable request.

\section*{Acknowledgments}
\noindent Y. T., W. G., and T. K. are supported by the National Science Foundation under Grant No. DMR-2100790 and the Gordon and Betty Moore Foundation through Grant GBMF11567. They also acknowledge the support and resources provided by the National High Magnetic Field Laboratory at Florida State University, which is supported by the National Science Foundation Cooperative Agreement No. DMR-1644779 and the state of Florida. M. T. acknowledges the support by the JSPS KAKENHI program under Grant No. JP20H01855. H. K. acknowledges the support by the JSPS KAKENHI program under Grant No. JP22H01403.

\section*{Author contributions}
\noindent W.G. designed and supervised the research and wrote the paper; Y.T. conducted the experiment; H.K. and Y.T. performed the numerical simulations; All authors participated in the result analysis and paper revision.

\section*{Competing interests}
\noindent The authors declare no competing interests.

\end{document}